# Two-dimensional conducting layer on SrTiO$_3$ surface induced by hydrogenation

Y. Takeuchi[1], R. Hobara[1], R. Akiyama[1*], A. Takayama[1#], S. Ichinokura[1], R. Yukawa[2], I. Matsuda[3] and S. Hasegawa[1]

[1]*Department of Physics, The University of Tokyo, 7-3-1 Hongo, Bunkyo-ku, Tokyo 113-0033, Japan*

[2]*Institute of Materials Structure Science, High Energy Accelerator Research Organization (KEK), Tsukuba, Ibaraki 305-0801, Japan*

[3]*Institute for Solid State Physics. The University of Tokyo, Kashiwa, Chiba 277-8581, Japan*

\# Present address: Department of Physics, Waseda University, 3-4-1 Ohkubo, Shinjuku-ku, Tokyo 169-8555

**Abstract**

We found that a surface state induced by hydrogenation on the surface of SrTiO$_3$(001) (STO) did not obey the rigid band model, which was confirmed by *in situ* electrical resistivity measurements in ultrahigh vacuum. With exposure of atomic hydrogen on the STO, a new surface state (H-induced donor state, HDS) appears within the bulk band gap (an in-gap state), which donates electrons thermally activated to the bulk conduction band, resulting in downward bending of the bulk bands beneath the surface. The doped electrons flow through the space-charge layer in two-dimensional manner parallel to the surface.

---

* akiyama@surface.phys.s.u-tokyo.ac.jp

The observed semiconducting behavior in the temperature dependence of electronic transport is explained by the thermal activation of carriers. The HDS and the bulk conduction band are non-rigid in energy position; they come closer with increasing the hydrogen adsorption. Eventually the HDS saturates its position around 88 meV below the bottom of the bulk conduction band. The sheet conductivity, accordingly, also saturates at 1.95±0.02 µS☐ with increasing hydrogen adsorption, corresponding to completion of the hydrogenation of the surface.

Surfaces and interfaces of perovskite-type oxides such as SrTiO$_3$ (STO), have attracted much attention since its electronic characteristics drastically change depending on the states of oxygen vacancies and surface termination. The surface is easily reconstructed to 1×1, 1×2, $\sqrt{5} \times \sqrt{5}$-R26.6°, and $\sqrt{13} \times \sqrt{13}$-R33.7° in accordance with the periodicity of oxygen vacancies on the surface which can be controlled by heat treatments [1-5]. Corresponding to such reconstructions, the electrical conductivity of surface also changes significantly [6]. Furthermore, the conducting surface layer of STO has been demonstrated to exhibit characteristics of two-dimensional electron gas (2DEG), 2D liquid (2DEL), and even 2D superconductivity [7-9]. The 2DEG observed at STO surface bears the aspect of the electrical conduction from a polaron phase to a metallic phase with changing the electron concentration [10,11].

We reported the appearance of 2DEL at the STO surface with hydrogenation as predicted by *ab initio* calculations [12], which was confirmed by angle-resolved photoemission spectroscopy (ARPES) [13,14]. According to this, with irradiating hydrogen atoms on TiO$_2$-terminated STO surface by cracking H$_2$ gas, a new metallic surface state appears around 60 meV below the Fermi level $E_\mathrm{F}$. Valence-band and core-level photoemission spectroscopy with *s* and *p*-polarized light also indicated that H atoms bond with oxygen atoms on STO surface; H-induced peaks in the photoemission spectra which lie across the valence band are attributed to the bonding state of $\sigma$(O-H) bond in the surface-normal direction [14]. This is consistent with the theoretical prediction that H atoms do not diffuse into the bulk STO crystal, but bond with O atoms on the $TiO_2$-terminated surface [12]. Although this surface state is expected to affect the transport properties because it is close to $E_F$, the electrical transport characteristics of the H-adsorbed STO (H-STO) surface have not been investigated in detail because the hydrogenated surface reacts easily with air at *ex situ* measurements.

In this study, *in situ* transport measurements of H-STO in ultrahigh vacuum (UHV) were performed systematically with non-doped STO substrates by means of a four-tip scanning tunnel microscope (STM). The results have revealed that a surface state appears within the bulk band gap by hydrogenation (a H-induced donor state, HDS), which donates electrons thermally activated to the bulk conduction band (BCB). In addition, the HDS becomes closer to the bottom of BCB in energy position with increasing adsorption of H atoms, and eventually is located at 88 meV below the BCB bottom. This change in energy position of HDS explains the semiconducting nature in temperature dependence of surface electrical conductivity. The HDS and bulk bands do not obey the rigid-band model during the irradiation of H atoms.

Transport measurements were carried out with a four-tip STM whose tungsten tips were controlled in position independently under scanning electron microscope (SEM) in the UHV environment with a base pressure in the range of $10^{-10}$ Torr. With this apparatus [15], transport properties such as the dimensionality and in-plane anisotropy in the electrical conduction can be analyzed [16]. The surface structure was checked by reflection high-energy electron diffraction (RHEED). Buffered-HF (BHF)-treated non-doped STO(001) wafers (SHINKOSHA Co., Ltd., Japan) were used as the samples (15 mm×3 mm×0.5 mm in size). They had steps along [100] direction. The BHF process makes STO(001) surface with the $TiO_2$-termination [17]. After installing the STO sample into the UHV chamber, a clean STO(001) surface was prepared by annealing at 650℃ in atmosphere of $3\times10^{-5}$ Torr of oxygen gas (99.999% purity) for 60 min. The STO was heated up by joule heating of a Si wafer (resistivity ~ 10 Ωcm) which was cramped on the backside of STO together. The annealing procedure in the oxygen atmosphere is well known to avoid oxygen vacancies on the surface [12,17]. The oxygen vacancies lower the electrical resistance because they donate electrons [18-22]. The 1×1 pattern was confirmed by RHEED after the annealing in oxygen, which was known to be the most stable structure of the clean $TiO_2$-terminated STO surface [23,24]. This 1×1 phase has few oxygen vacancies on the surface [25]. The contact resistance between the W STM tip and the sample surface was high (> 1 TΩ), meaning that the bare STO surface is not metallic.

Then, the sample was set in front of a heated W filament for cracking $H_2$ molecules in the atmosphere of hydrogen gas (99.99% purity) to adsorb atomic H on the STO(001) surface. The distance from the sample to the filament was fixed at 15 cm. The hydrogenation was performed under the $H_2$ gas pressure of $2\times10^{-5}$ Torr with the STO substrate kept at room temperature. The exposure time was changed under control to investigate the dependence of resistivity on H-adsorption amounts. After irradiation of hydrogen for 50 kilo-Langmuir (kL), the 1×1 streaks in RHEED pattern became sharper than those of the clean surface, as shown in Fig. 1 (a) (left; patterns, and right; line profiles of streaks ). Figure 1(c) (bottom) shows the change in the width of the (01) and (0$\bar{1}$) streaks in RHEED as a function of the amount of hydrogen irradiation. The streaks become sharper especially over 30 kL exposure of hydrogen. One possible reason of this change in RHEED patterns is that the crystallinity and flatness of the surface was improved, and the other possibility is that the charge-up effect on the insulating STO substrate was reduced after the hydrogen adsorption. The irradiated hydrogen atoms are known to adsorb by bonding with O atoms on the surface [13], as shown in Fig, 1(b) (bottom). Another previous report indicated that an O atom which bonds with a neighbor Ti atom vibrates along perpendicular to the Ti-O bond [26]. With hydrogen irradiation, such vibration may be suppressed due to emergence of the O-H bond.

The contact resistance between the W STM tip and the sample surface decreased significantly from over 1 TΩ to 1 MΩ by the hydrogen adsorption. By using the square four-point probe (4PP) method with the tip configuration as shown in Fig. 1(b) (top) at room temperature, the conductivity increased with the hydrogen irradiation and saturates at 1.95±0.02 μS□ over 50 kL irradiation as shown in Fig. 1(c) (top). Such an increase of conductivity was also observed in the previous study [13]. This is known by photoemission spectroscopy to be due to electron doping which causes the bulk-band downward bending near the surface. However, the magnitude of the saturated conductivity in the present study is much smaller than that in the previous one (440 μS□) [13]. This is because in the previous study [13] a highly-Nb-doped STO substrate was used to prevent charge-up in ARPES measurements so that $E_F$ crossed the BCB by hydrogenation while in the present study we used a non-doped STO substrate having $E_F$ well below the bottom of BCB even with the hydrogenation as described below. Thus, the concentration of electrons accumulated in the 2DEL beneath the surface of the previous study is much more than that of the present study.

The annealing in oxygen was performed after the hydrogenation to remove the hydrogen atoms from the surface, and then the hydrogenation was done again on the same sample. This procedure was defined as "cycle" in Fig. 1(c) (top). The change in conductivity during the hydrogenation was almost reproduced well by repeating the cycles of oxygen-annealing and hydrogenation. This suggests that hydrogen desorbed reversibly from the surface by the oxygen annealing.

The carrier type of the H-STO surface was found to be $n$-type from the rectify characteristic between the W STM tip and H-STO as shown in Figs. 2(a) and (b); the current flows from the tip to the sample when the positive bias voltage was applied to the tip.

In Fig. 2(c), the four-point-probe sheet conductivity at different exposures of hydrogen is plotted as a function of the inverse temperature. The conductivity can be fitted by the Arrhenius equation as follows,

$$\sigma = C \exp\left(-\frac{E_a}{k_B T}\right), \tag{1}$$

where $C$, $E_a$, and $k_B$ are a fitting coefficient, an activation energy, and Boltzmann constant, respectively. We can see from this result that the activation energy depends on the amount of hydrogenation; the energy drastically decreases with increasing the hydrogen exposure. This means that the energy gap between the HDS and the BCB bottom, which corresponds to the activation energy $E_a$, decreases with the hydrogenation. In addition, since the values of $E_a$ are much smaller than the band gap of STO (3.2 eV), the HDS is an in-gap state, located within the band gap.

The schematic of the band diagram based on these facts is shown in Fig. 3. The BCB is bent downward beneath the surface to create a surface space charge layer (SCL). This is because electrons are thermally activated from the HDS to be doped into the BCB. Such electrons are transported along the SCL to reduce the sheet resistivity. Since any anisotropy in conductivity was not observed by the square 4PP method as described later, the electrical conduction is due to the 2DEL at the SCL [8].

Contrary to the previous works [13,14] in which highly doped STO substrates were used, the present work employed the non-doped STO to measure the lateral electrical transport near the surface in detail. Thus, the initial $E_F$ position in the present work is located around the middle of the band gap of STO while it is very close to the bottom of the BCB in the previous studies. Despite of such differences, the mechanism of the band bending is basically the same; the band is bent downward with the electron doping to the BCB from the HDS by hydrogenation. Since surface states usually move together with the bulk bands during band bending, the energy difference between the HDS and the bottom of BCB is expected to be constant irrespective of the degree of band bending. In other words, the surface energy levels with respect to the bulk states are defined by the structure (rigid band model [27]). However, intriguingly, the activation energy changes during hydrogenation in the present system, indicating that the HDS does not obey the rigid band model. This is because some structural changes occur during the hydrogenation, which modulates the energy difference between the HDS and the bulk bands. In fact, the theory predicts that the hydrogenation makes the surface atoms shifted with the coverage of hydrogen [12]. According to this theory, oxygen atoms are pulled out of the surface by bonding with the hydrogen atoms. As shown in Fig. 3, the BCB is bent downward to be closer to the HDS (*i.e.* $E_a$ decreases) and it finally saturates at 88 meV above the HDS (Figs. 1(c) and 2(c)). This is because the HDS is one of the in-gap states (IGSs) as reported in many papers [13,14,28-31]; various kinds of IGSs have been reported so far around $E_F$ after irradiation of H atoms [32,33]. Actually, despite another oxide, ZnO, it also shows the downward bulk band bending with hydrogenation, and the photoemission intensity of the H-O bonds reaches saturation because the adsorption probability of H atom is lowering with increasing H exposures [34]. In this report, such hydrogenation dependence of the degree of the band bending was observed by ARPES. Analogically we infer that the similar situation is realized in our case of STO.

Here, we should infer another aspect of the observed change in sheet conductivity as a function of H exposure (Fig. 1(c)). With a simple model of Langmuir-type adsorption model (without desorption) of H atoms on the STO surface, the coverage of adsorbed H, and therefore the density of states of HDS, are proportional to $(1 - e^{-\alpha D})$ where $D$ is the H exposure and $\alpha$ is a coefficient, respectively. This relation also should be applied to the carrier density activated in the BCB, and therefore proportional to the sheet conductivity. However, the change in sheet conductivity in Fig. 1(c) does not obey this expectation, meaning that the main reason for the conductivity increase with the H exposure is not due to the increase of the density of states of HDS, rather due to the decrease in the activation energy as mentioned above.

In addition, we measured in-plane anisotropy in transport at H-STO with the square 4PP configuration shown in Fig. 4(a) and Fig. 1(b) (top). However, the samples measured in this study did not show any anisotropy as shown in Fig. 4(b), which is different from the results of surface-state transport on silicon surfaces [35]. Since the present result means that the carrier scattering at steps on the surface does not affect the electrical conduction, we can say that the electrical conduction in the present case is not at the topmost layer of the surface (surface-state transport), but at the SCL beneath the surface. This indicates that the estimated $E_a$ does not denote the activation energy of the hopping between HDSs, because the hopping generally occurs on the topmost layer of the surface where it should be easily influenced by the anisotropic step configurations. The HDS is a localized state due to H-O bonding, which does not directly contribute to the conduction parallel to the surface, while the electrons donated from the HDS to the SCL are transported at the SCL parallel to the surface. The thermal activation from the HDS to the BCB band results in the semiconducting temperature dependence of resistivity shown in Fig. 2(c).

The dimensionality of the electrical transport can be distinguished by the 4PP method with different probe distances. When the sample is semi-infinite three-dimensional bulk, the observed resistance $R$ is inversely proportional to the probe distance $d$. On the other hand, when the system is semi-infinite two-dimensional sheet, $R$ does not depend on $d$ as described the following equation [15],

$$R = \frac{\ln(1+\frac{\sigma_x}{\sigma_y})}{2\pi\sqrt{\sigma_x\sigma_y}}, \quad (2)$$

where $\sigma_{x,y}$ is the sheet conductivity along $x$ or $y$ direction. If there is no anisotropic conductivity like the present case, Eq. (2) can be simply written as ($\sigma_x = \sigma_y \equiv \sigma_{2D}$)

$$R = \frac{\ln 2}{2\pi\sigma_{2D}}. \quad (3)$$

In our result, as shown in Fig. 4(c) the resistance is independent of the probe distance $d$, and thus the two-dimensional transport is realized for the H-STO. The sheet conductivity $\sigma_{2D}$ was estimated to be $0.36 \pm 0.06$ μS☐ by Eq. (2). The sheet conductivity contributed by the SCL, $\sigma_{SCL}$ can be calculated by solving Poisson's equation using bulk parameters such as electron and hole mobility, dielectric constant, and band gap [36,37]. In the present case, by considering the surface $E_F$ position to be 88 meV below the BCB bottom, the $\sigma_{SCL}$ in STO was estimated to be 1.33 μS□, which is on the same order of the observed sheet conductivity. This means that the transport dominantly comes from the two-dimensional SCL in this system.

In summary, the sheet conductivity increased with adsorption of atomic hydrogen on the STO(001) surface, and it saturated at 1.95 ± 0.02 μS□ with the saturation adsorption. The HDS is created by the hydrogenation and donates electrons into the STO substrate, resulting in downward bending of the bulk bands to create the SCL beneath the surface. The SCL is responsible for the two-dimensional electrical conduction parallel to the surface. We have found from the temperature dependence of the conductivity that electrons are thermally activated from the HDS to the SCL and that the activation energy decreases with the amount of hydrogen adsorbed and eventually saturates at the minimum value of 88 meV. This implies that the HDS shows non-rigid-band nature and located within the band bulk gap, so-called an IGS. These results on the surface conductivity at STO with hydrogenation shed light on correlation of the atomic structures with physical properties of perovskite oxides and point to a possibility for utilizing it.

**Acknowledgment**

This work was partially supported by a Grant-in-Aid for Scientific Research (A) (KAKENHI No. JP16H02108), a Grant-in-Aid for Young Scientists (B) (KAKENHI No. 26870086), a Grant-in-Aid for Challenging Exploratory Research (KAKENHI No. 15K13358), Innovative Areas "Topological Materials Science" (KAKENHI No. JP16H00983, JP15K21717), and "Molecular Architectonics" (KAKENHI No. 25110010) from Japan Society for the Promotion of Science.**Reference**

[1] H. Tanaka, T. Matsumoto, T. Kawai, S. Kawai, *et al.*, Jpn. J. Appl. Phys. **1**, 1405 (1993).

[2] E. Heifets, *et al*., Phys. Rev. B **75**, 115417 (2007).

[3] S. Ogawa, *et al*., Phys. Rev. B **96**, 085303 (2017).

[4] R. Shimizu, K. Iwaya, T. Ohsawa, S. Shiraki, T. Hasegawa, T. Hashizume, and T. Hitosugi, Appl. Phys. Lett. **100**, 263106 (2012).

[5] M. Naito and H. Sato, Phys. C Supercond. **229**, 1 (1994).
[6] A. Spinelli, *et al,* Phys. Rev. B **81**, 155110(2010).

[7] A. F. Santander-Syo, O. Copie, *et. al*., nature **469**, 189 (2011).

[8] Z. Wang *et al*., Nat. Mater. **15**, 835 (2016).

[9] M. Kim, *et al*., Phys. Rev. Lett. **107**,106801 (2011).

[10] A. F. Santander-Syro *et al.*, nature **469**, 189 (2011).

[11] C. Chen, J. Avila, E. Frantzeskakis, A. Levy and M. C. Asensio, Nat. Commun. **6**, 8585 (2016).

[12] F. Lin, S. Wang, F. Zheng, G. Zhou, J. Wu, B.-L. Gu and W. Duan, Phys. Rev. B **79**, 035311 (2009).

[13] M. D'Angelo, R. Yukawa, K. Ozawa, S. Yamamoto, T. Hirahara, S. Hasegawa, M. G. Silly, f. Sirotti, and I. Matsuda, Phys. Rev. Lett. **108**, 116802 (2012).

[14] R. Yukawa, S. Yamamoto, K. Ozawa, M. D'Angelo, M. Ogawa, M. G. Silly, F.14

**Figure captions**

FIG 1 (a) (left) RHEED patterns of STO(001)-1×1 surface and H-STO(001)-1×1 (50 kL exposure of hydrogen). (right) Intensity profiles of (0$\bar{1}$) and (01) streaks along red dashed lines in RHEED patterns at 0 L, 30 kL and 50 kL exposure of hydrogen. These profiles were fitted well with Gaussian functions. (b) (top) SEM image of the *in-situ* square four-point probe measurement configuration, and (bottom) a schematic picture of the atomic arrangement of H-STO surface. (c) (top) The dependence of sheet conductivity measured at room temperature upon the H-exposure. A red dashed curve is a guide for eyes. The resistance was measured by the square four-point probe method with fixing the probe distance (150 μm). (bottom) The FWHM of (01) and (0$\bar{1}$) streaks in RHEED pattern, obtained by the Gaussian fitting (see (a)), as a function of the H-exposure.

FIG 2 (a) A schematic of one-probe method, and (b) *I-V* curve with this configuration. (c) The four-point-probe sheet conductivity as a function of inverse temperature at different exposures of hydrogen. The probe distance in the square-4PP measurement was 150 μm. The lines represent the fitting by the Arrhenius law Eq. (1). The activation energies were estimated by the fitting of red circles, green squares and blue rectangles to be 88 ± 2 meV, 118 ± 8 meV and 660 ± 43 meV, respectively.

FIG 3 Schematic band diagram at the hydrogenation. $E_a$ is the activation energy estimated in Fig. 2 (c). The bulk bands are bent downward near the STO surface by electron doping thermally activated from the HDS. The electrons flow through the SCL parallel to the surface.

FIG 4 (a) A schematic of the square 4PP method, and (b) *I-V* curve with the method. $R\perp$ step : red dots (*R*//step : blue dots) means that the current flows between probe #1(#1) and #4(#2), which corresponds to the step-perpendicular (step-parallel) direction, and the voltage was measured between probe #2(#3) and #3(#4). In both configurations, the slope of *I-V* curve is almost the same each other. (c) Probe-distance *d* dependence of resistance *R* measured by the square 4PP method. The yellow dashed line represents the behavior of three-dimensional resistance ($R \propto d^{-1}$), whereas the red and blue lines indicate that of the two-dimensional one ($R \propto d^0$) in the cases of $R\perp$step and *R*//step, respectively.

**Figures**

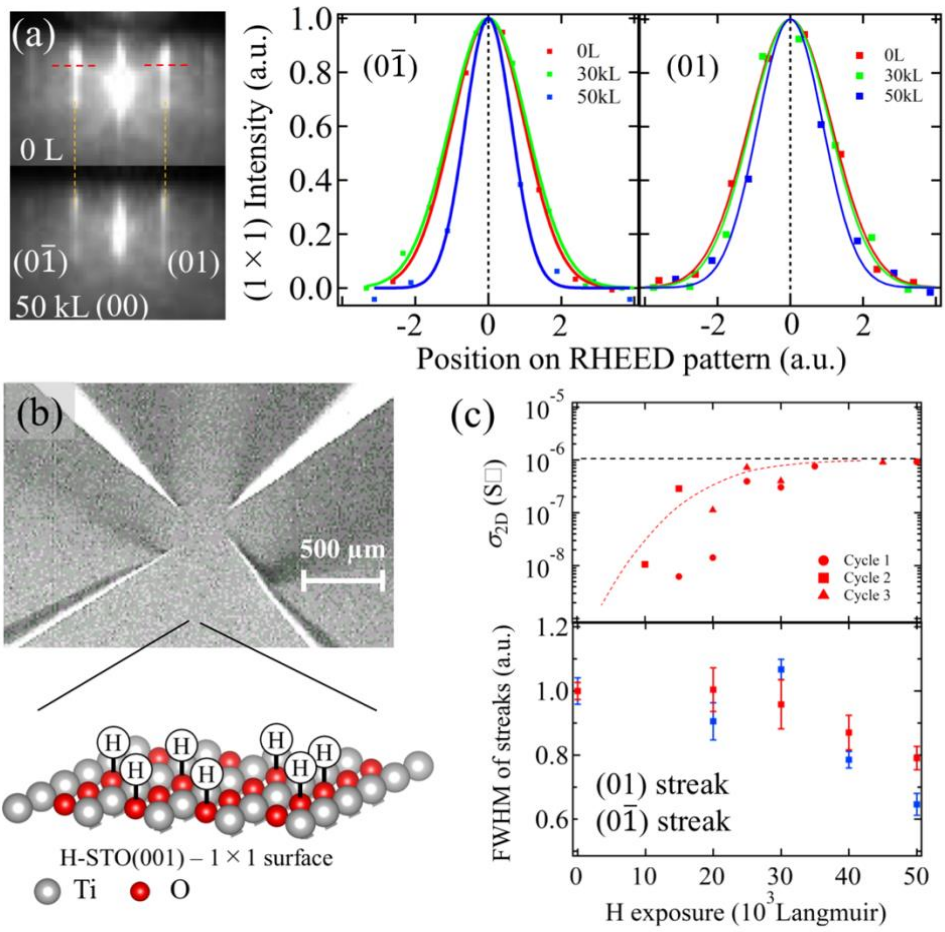

Figure 1

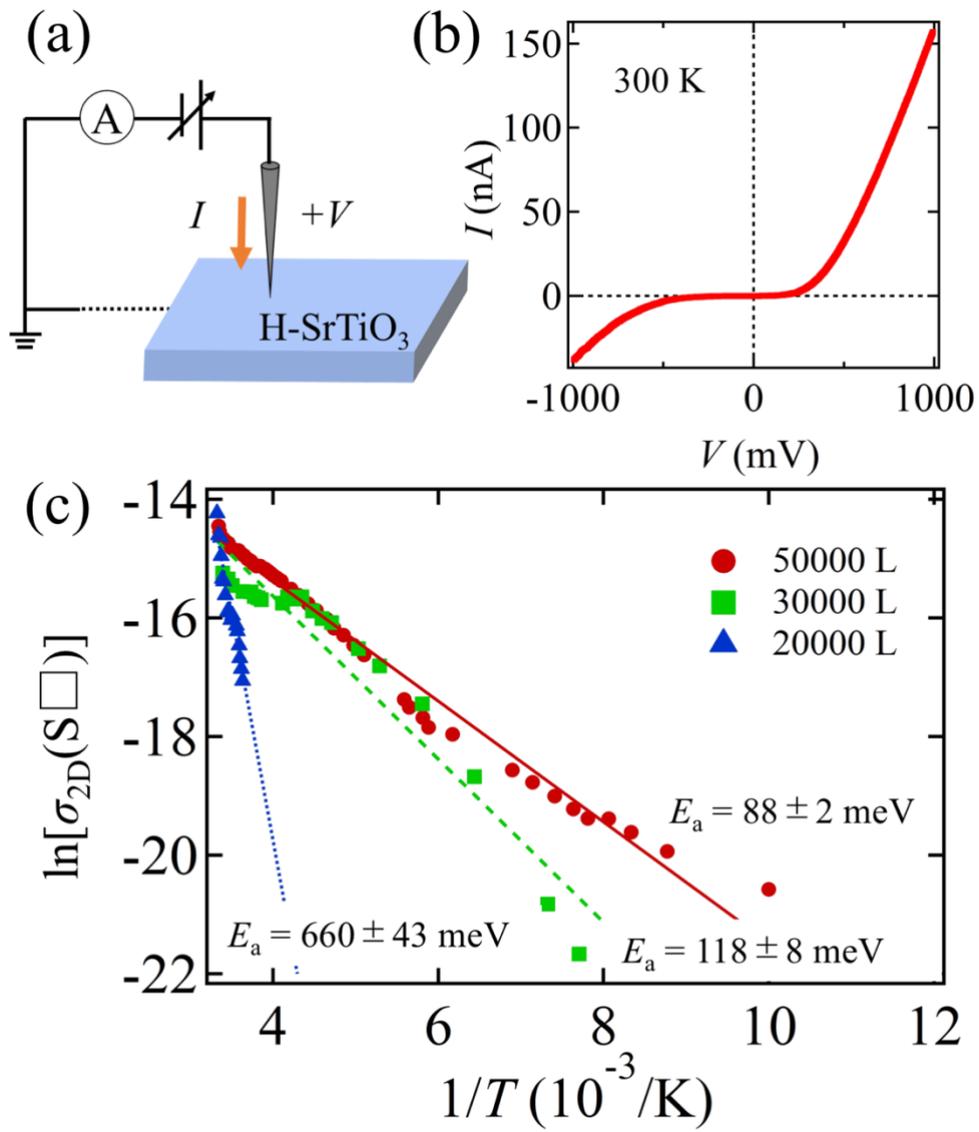

Figure 2

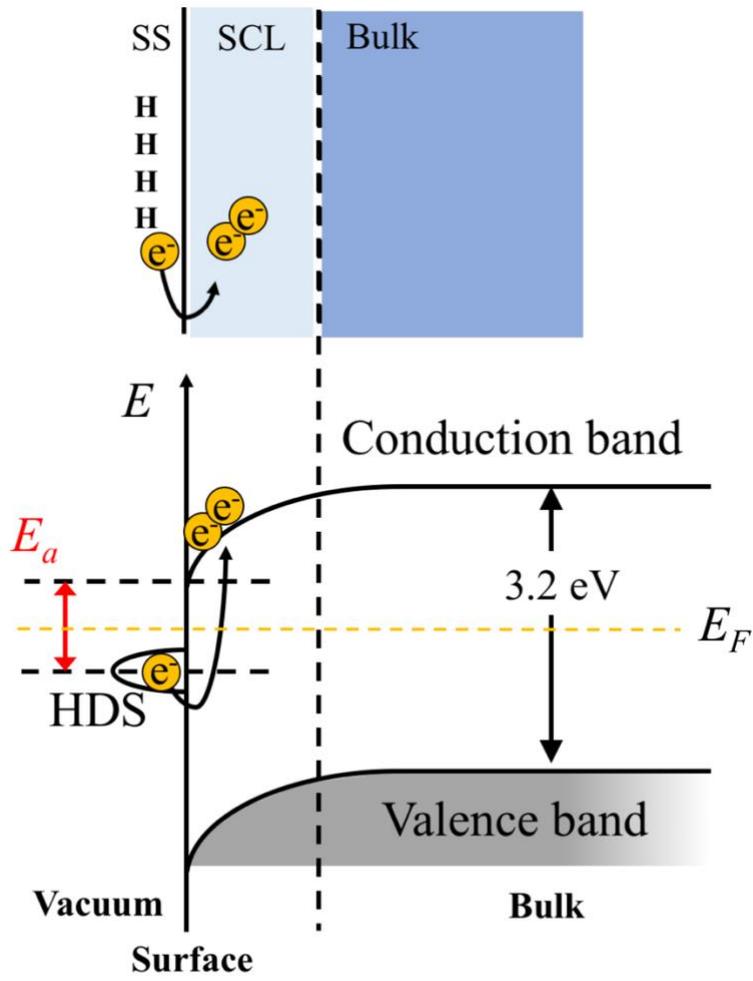

Figure 3

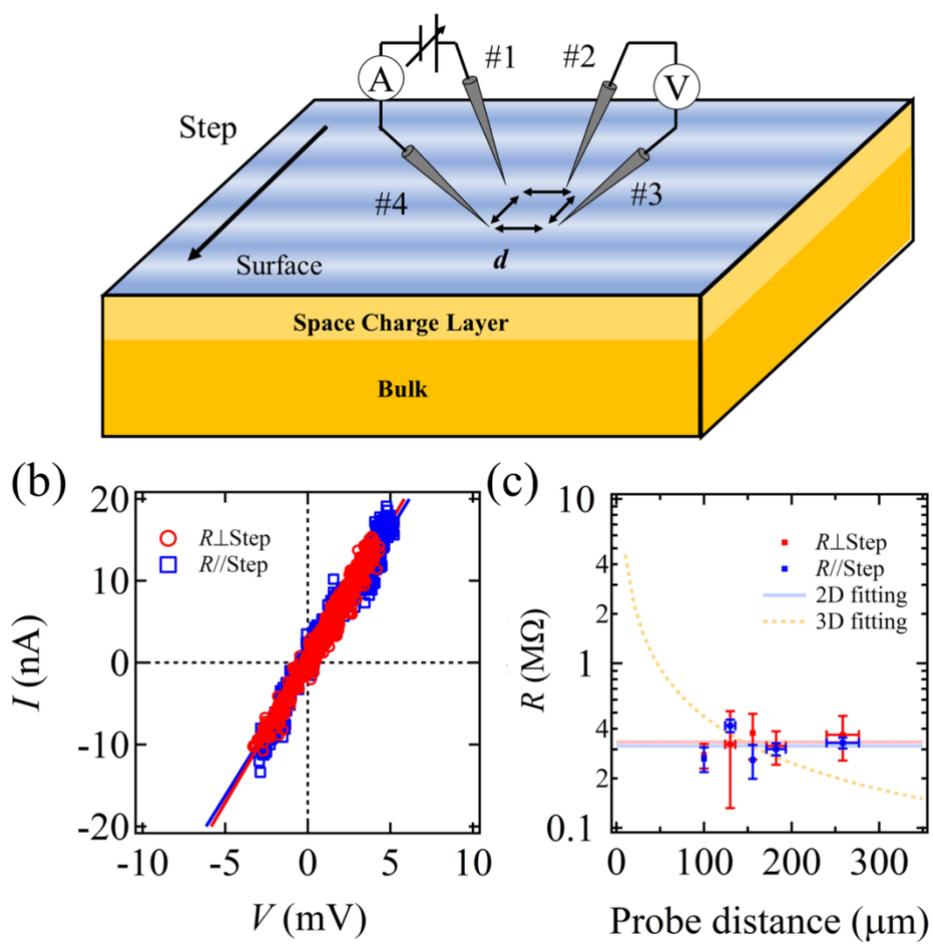

Figure 4